\documentclass[aps,notitlepage,a4paper,superscriptaddress,12pt]{article}

\usepackage[utf8]{inputenc}
\usepackage[dvips]{color}
\usepackage[toc]{appendix}
\usepackage[hang,flushmargin]{footmisc} 
\usepackage{titlesec,titletoc,ntheorem-hyper,abstract,latexsym,microtype,booktabs,hyperref,amsmath,cleveref,fancyhdr,wrapfig,array,amsfonts,amssymb,geometry,appendix,cite,float,lipsum,setspace,xcolor}
\usepackage{soul}
\usepackage[affil-it]{authblk} 
\usepackage{dsfont}
\usepackage{amsmath, amssymb, graphics, setspace}
\newcommand{\lambdabar}{{\mkern0.75mu\mathchar '26\mkern -9.75mu\lambda}}

\newcommand{\be}{\begin{equation}}
\newcommand{\ee}{\end{equation}}
\newcommand{\bea}{\begin{eqnarray}}
\newcommand{\eea}{\end{eqnarray}}
\newcommand\blfootnote[1]{%
  \begingroup
  \renewcommand\thefootnote{}\footnote{#1}%
  \addtocounter{footnote}{-1}%
  \endgroup
}

\usepackage{caption}
\usepackage{subcaption}
\usepackage{amsmath}
\usepackage{amsfonts}
\usepackage{amssymb}
\usepackage{mathtools}
\numberwithin{equation}{section}
\usepackage{titlesec}
\usepackage{sectsty}
\sectionfont{\centering}
\usepackage{framed}
\allowdisplaybreaks
\usepackage[nottoc]{tocbibind}
\usepackage[free-standing-units]{siunitx}
\usepackage{helvet}
\usepackage{url}
\usepackage{titletoc}
\usepackage{blindtext}
\usepackage[gen]{eurosym}

\title{Refractive index profiles for a $\mathcal{PT}$-symmetric optical structure} 

 \author{Bijan Bagchi$^{1,\dagger,*}$, Rahul Ghosh$^{2,3,\dagger}$ and Sauvik Sen$^{3,\dagger}$}
\affil{1. Brainware University, Barasat, Kolkata 700125, West Bengal, India\\\vspace{1em}2. The Dynamics Lab, Department of Chemistry,\\ Indian Institute of Technology Delhi, New Delhi 110016, India\\
\vspace{1em}3. Department of Physics, Shiv Nadar Institution of Eminence,\\ Gautam Buddha Nagar, Uttar Pradesh 203207, India}

\begin{document}
\maketitle
\begin{abstract}
 By mapping the scalar Helmholtz equation (SHE) to the Sch\"{r}odinger form  we investigate the behaviour of $\mathcal{PT}$ optical structure when the refractive index distribution $n$ admits variation in the longitudinal direction only.  Interpreting the Sch\"{r}odinger equation in terms of a superpotential we determine the supersymmetric partners for $n$. We also obtain new analytical solutions for the refractive index profiles and provide graphical illustrations for them.\\
  

\end{abstract}

\blfootnote{e-mail : 1. bbagchi123@gmail.com, 2. rg928@snu.edu.in, 3. sauviksen.physics@gmail.com}
\blfootnote{$*$ : corresponding author}
\blfootnote{$\dagger$ : These authors contributed equally to this work.}



\section{\label{sec:level1}Introduction}

The scalar Helmholtz equation (SHE) is an elliptic differential equation which corresponds to the eigenvalue problem of the Laplacian operator \cite{vla}. For a propagating electric field $\mathcal{E}$ along the $z$-axis (i.e. in the longitudinal direction), it takes the form  

\begin{equation}\label{SHE}
    \left ( \partial_z^2  +  \partial_x^2  +k^2 \right ) \mathcal{E}  =0, \
\end{equation}
where $k = \frac{2\pi}{\lambda}$ is the transverse wave number. On factorizing the operator acting on $\mathcal{E}$ we can choose, without loss of generality, for one of the factors to satisfy

\begin{equation}\label{factorized}
    \left (i \partial_z  + \sqrt{\partial_x^2 + k^2} \right ) \mathcal{E} = 0 
\end{equation}
On substituting $\mathcal{E} = e^{ik z} \phi (x, z)$ we obtain

\begin{equation}\label{SE1}
    \frac{i}{c}\partial_t \phi - k \phi +k \left (1 + \frac{\partial_x^2}{k^2} \right )^{\frac{1}{2}} \phi =0 
\end{equation}
where we have identified the variable $z$ with $ct$. Using the standard formula for the binomial expansion $(1+z)^\frac{1}{2} = 1 + \frac{1}{2}z -\frac{1}{8} z^2 +...$ we can express

\begin{equation}\label{Taylor}
\left (1 + \frac{\partial_x^2}{k^2} \right )^{\frac{1}{2}} =  1 + \frac{1}{2}\frac{\partial_x^2}{k^2} -\frac{1}{8} \frac{\partial_x^4}{k^4} +...
\end{equation}
thereby obtaining from the form 

\begin{equation}\label{SE2}
    \frac{i}{c}\partial_t \phi = -\frac{1}{2} \left (\frac{\partial_x^2}{k} \right )\phi + \frac{1}{8} \left (\frac{\partial_x^4}{k^3} \right )\phi
\end{equation}
where we retained terms up to fourth order in $\partial_x$. Setting $\hat{p} =-i\hbar \partial_x$, we arrive at 

\begin{equation}\label{SE3}
    i\hbar\partial_t \phi = \frac{c}{2k}\frac{\hat{p}^2}{\hbar} \phi + \frac{c}{8k^3} \frac{\hat{p}^4}{\hbar^3}\phi
\end{equation}

The representation (\ref{SE3}) affords comparison with a fourth-order quantum nonlinear Schr\"odinger equation \cite{das1, das2} 

\begin{equation}\label{SE4}
    i\hbar\partial_t \psi = \frac{\hat {p}^2}{2m} \psi  +  \frac{\tau \hat {p}^4}{3m} \psi
\end{equation}
where a modification \cite{conti1, conti2} under the influence of a generalized uncertainty principle (see, for example, \cite{akempf, gomes, bagchi, quesne, dey, mandal, hussin, hossenfelder, lake} and refereces therein) has been taken into account. With $\lambdabar = \frac{\lambda}{2\pi}$ and $k=\frac{1}{\lambdabar}$, we readily find the following set of consistency conditions

\begin{equation}\label{compare1}
\frac{1}{2m}=\frac{c}{2\hbar k} \Rightarrow mc=\frac{\hbar}{\lambdabar}, \quad k =\frac{mc}{\hbar}
\end{equation}
and

\begin{equation} \label{compare2}
\frac{\tau}{3m} = \frac{c}{8\hbar^3 k^3} = \frac{c}{8m^3 c^3} \Rightarrow \tau = \frac{3}{8} \frac{1}{m^2c^2}
\end{equation}
In other words we have for $\tau$ the result
$\tau = \frac{3}{8} \left (\frac{\lambdabar}{\hbar} \right )^2$. 
There are various ways to estimate $\tau$ depending upon the theory at hand. It can be related to the dimensionless quantity $\tau_0$ by defining $\tau_0 = \frac{\hbar c^3}{G} \tau$.
In terms of the Planck mass, 
$\tau_0$ translates to $\tau_0 = m_P^2 c^2\tau$,
and we are led to the connection

\begin{equation} \label{tau_0}
\tau_0 = m_P^2 c^2 \left (\frac{3}{8} \frac{1}{m^2c^2} \right ) = \frac{3}{8} \frac{m_P^2}{m^2}
\end{equation}
It is evident that the above estimate of $\tau_0$ is dependent on the mass of the subatomic particle under consideration. 
For an electron mass of $\approx 9.1 \times 10^{-31}$ kg, we find  $\tau_0$ to be of the order of  $\sim 10^{44}$ which is about $O(10^{11})$ lower than what is obtained from the photon. Our estimate of $\tau_0$ is in accordance with the idea of an intermediate length between the Planck scale and the electroweak scale given by \cite{das1}. Quantum optical tenchniques are being incorporated to design experiments to study Planck-scale physics \cite{Piko}. Generalized uncertainty principle plays an important role in designing such optomechanical systems \cite{Sen}.

This paper is organized as follows: In section 2 we explore the formal equivalence of quantum mechanics and optics by disregarding the paraxial approximation where the refractive index has only a transverse ($x$) component.  In section 3, we consider an optical
periodic structure that addresses a distribution which varies only in the longitudinal ($z$) direction. It enables us to set up a scheme in which the SHE is presented with a supersymmetric structure. As a result, we are able to derive new analytical forms for the complex periodic partner of the refractive index distribution. This section also addresses  the question of determining closed form solutions of the parity-time symmetric periodic structure of the refractive index, where the parity operator $\mathcal{P}$ is defined by the operations $(i, x, p) \rightarrow (i, -x, -p)$ and time reversal operator $\mathcal{T}$ by the ones $\rightarrow (-i, x, -p)$. It is of interest to mention here that in recent times the idea of $\mathcal{PT}$ has found relevance in the artificial construction of optical structures with balanced gain and loss \cite{zya}. In section 4, we make a summary of our results.

\section{Non-paraxial scalar Helmholtz equation and Schr\"{o}dinger equation equaivalence}

We focus on the paraxial approximation
in which the scalar wave equation could be shown equivalent to an analogue of Schr\"{o}dinger equation of a two-dimensional harmonic oscillator \cite{ste}. The work of Lin et al \cite{lin} aroused much interest after it tried to explore a nonparaxial model wherein the consequences of exploring $\mathcal{PT}$ were examined to achieve  unidirectional invisiblity at the exceptional points with the refractive index distribution being entirely longitudinally directed. Their idea was taken up by Jones \cite{jones1} (see also \cite{jones2}) to derive analytical conditions for a $\mathcal{PT}$ optical structure. In particular, in this work, he enquired into the question of how a complex refractive index offers an understanding of unidirectional invisibility in $\mathcal{PT}$-symmetric systems. The basic point was to exploit only the z-variation of the refractive index rather than the typical paraxial exercise where the variation of $n$ is taken in the transverse direction\footnote{Complex, transversely distributed refractive index which is inherently $\mathcal{PT}$-symmetric and playing the role of an optical potential has been widely studied \cite{mak}. The corresponding electric-field envelope $\mathcal{E}$ then obeys the paraxial equation of diffraction. For a study of the general class of index profiles see \cite{mos}.}. In the  setting of \cite{lin}, the SHE acquires a form similar to the one-dimensional time-independent Schr\"{o}dinger equation but endowed with a spatial z variable 

\begin{equation}  \label{nonparaxHelm}
 \frac{d^2 \mathcal{E}}{dz^2} + k^2 \left (\frac{n}{n_0} \right )^2 \mathcal{E} = 0  
\end{equation}

Let us look at the corresponding stationary Schr\"{o}dinger equation governing a quantum particle influenced by a complex optical potential $V^{(+)}(z)$ in the following dimensionless form  ($\hbar = 2m = 1$). Here the 'plus' in the superscript of $V$ is a reference to the positive supersymmetric partner potential, details of which have been discussed in the subsequent sections.
\begin{gather} \label{00schrodingerE}
    -\frac{d^2\psi}{dz^2}+ \Big(V^{(+)}(z) -\varepsilon \Big)\psi = 0
\end{gather}
where $\varepsilon \in \Re$ is an incident energy scale. Comparison of (\ref{nonparaxHelm}) and (\ref{00schrodingerE}) shows that the role of the wave function $\psi(z)$ is analogous to the electric field amplitude $\mathcal{E}$ \cite{longhi} while the object $\frac{d^2 }{dz^2} + k^2 \left (\frac{n}{n_0} \right )^2$ transforms like the Schr\"{o}dinger operator and pointing to the connection 
\begin{align} \label{Vandnrelation}
   V^{(+)} = \varepsilon - k^2 \left(\frac{n}{n_0}\right)^2 
\end{align}

In this work, we attempt to interpret (\ref{Vandnrelation}) in the framework of supersymmetric quantum mechanics (SQM) and make an assumption that the index
distribution admits a lowest energy bound state specified with a propagation eigenvalue \cite{miri1}. The study of supersymmetric optical structures was carried out in some detail in \cite{miri2} to establish a relationship between two wave optical structures. It is well known that the formalism of SQM is primed to reveal new types of associated spectral problems through the existence of superpartners by utilising the so-called factorization method \cite{miel2004, dong} or equivalently making use of the intertwining conditions \cite{iof2012, fer} which emerge as a set of consistency relations. As we shall presently see, the Helmholtz equation, by enforcing the above analogy, also points to a new type of partner potential that is tied up to $V^{(+)}(z)$.
 
 A few words on SQM are relevant here. Its basic formalism \cite{jun1996, bag2000, kha2001, fer2010} involves a pair of odd operators $Q, Q^\dagger$ that generate the Hamiltonian in the form of an anticommutation relation
$\mathcal{H} = \{Q, Q^\dagger \}$. These operators obey the closed graded algebra $(Q)^2 = 0 = (Q^\dagger)^2, \quad
[Q, \mathcal{H}] = 0 = [Q^\dagger, \mathcal{H}]$
and could be represented in terms of operators $\mathcal{O}$ and $\mathcal{O}^\dagger$ such that $Q = \mathcal{O} \otimes \sigma_-, \quad Q^\dagger = \quad \mathcal{O}^\dagger \otimes \sigma_+$, 
where the quantities $\sigma_{\pm}$ denote $\sigma_{\pm} = \frac{1}{2}(\sigma_1 \pm i \sigma_2)$, with $\sigma_1$ and $\sigma_2$ are the usual Pauli matrices.  Taking a first-order differential realization of $\mathcal{O},\mathcal{O}^\dagger$  

\begin{equation}  \label{operatorsO}
    \mathcal{O} = \partial + \mathcal{W}(z), \quad \mathcal{O}^\dagger = -\partial + \mathcal{W}(z)
\end{equation}
where $\partial \equiv \frac{d}{dz}$ and $\mathcal{W}(z)$ is the superpotential of the system, we can project $Q$ and $Q^\dagger$ in the matrix forms

\begin{equation}
   Q = \left( \begin{array}{cc} 0 & 0  \\ \partial + \mathcal{W}(z)  & 0  \end{array} \right), \quad Q^\dagger =  \left( \begin{array}{cc} 0 &  -\partial + \mathcal{W}(z) \\0 & 0  \end{array} \right)
\end{equation}
 The Hamiltonian $\mathcal{H}$ is thus rendered diagonal 
whose two elements are $H^{(+,-)}$ given by the components $\mathcal{H}^{(+)} = \mathcal{O} \mathcal{O}^\dagger = -\partial^2 + V^{(+)} (z) - \Lambda, \quad 
 \mathcal{H}^{(-)} = \mathcal{O}^\dagger \mathcal{O}  = -\partial^2 + V^{(-)} (z) - \Lambda$,
both of which are in the typical Sch\"{r}odinger form defined at some cut-off energy value $\Lambda \in \Re$. In terms of $\mathcal{W}(z)$, the SUSY partner potentials $V^{(+,-)}$ can be projected as

\begin{equation}{\label{V+-}}
    V^{(+,-)} = \mathcal{W}^2(z) \pm \mathcal{W}'(z) + \Lambda
\end{equation}
where the prime represents a derivative with respect to $x$ and we identify $V^{(+)}$ with Schr\"{o}dinger potential in (\ref{00schrodingerE}).

For unbroken $SUSY$, the partner Hamiltonians have nonnegative energy eigenvalues with the ground state wavefunction $\psi_0 (z)$ being non-degenerate which we associate with the component $\mathcal{H}^{(-)}$; other energy levels of the partner Hamiltonians are positive and degenerate. The form of the ground state wavefunction can be obtained by solving $\mathcal{O} \psi_0 (z) = 0$ which means

\begin{equation}
\psi_0 (z) \propto \exp \left ( -\int^z \mathcal{W}(t) dt \right )
\end{equation}
reflecting that the normalizability of $\psi_0$ restricts the superpotential to obey $\int^x W(t) dt > 0$ as $z \rightarrow \infty$. 
The double-degeneracy of the spectrum is guided by the intertwining relationships of 
$\mathcal{O} \mathcal{H}^{(-)} = \mathcal{H}^{(+)} \mathcal{O}, \quad  \mathcal{H}^{(-)} \mathcal{O}^\dagger= \mathcal{O}^\dagger\mathcal{H}^{(+)} $ and furnish the isospectral connections between $\mathcal{H}^{(+)}$ and $\mathcal{H}^{(-)}$. Interest in SQM has been very recently revived in connection with its experimental realization in a trapped ion
quantum simulator as was reported in \cite{cai}.

\section{Partner refractive index profiles}

Returning to the non-paraxial equation (\ref{nonparaxHelm}), let us identify a pair of partner potentials $V^{(+,-)}$ in the following Riccati form which is induced by a  superpotenial $W$ \cite{wolf} with some cut-off energy value $\lambda \in \Re$ akin to (\ref{V+-})
\begin{gather} \label{Vto W}
    V^{(+,-)} = W^2(z) \pm W'(z) + \lambda
\end{gather}
 (\ref{Vandnrelation}) implies the corresponding relations for the index profiles  
\begin{align} \label{n2=W2W'}
    k^2 \left(\frac{n^{(+,-)}}{n_0}\right)^2 = \varepsilon -\lambda -(W^2(z) \pm W'(z)) 
\end{align}
where we fix the $W$ by a complex decomposition \cite{and1999, bagchi2001}
 \begin{align} \label{W(z)fg}
     W(z) = f(z)+i g(z)
 \end{align} 
It gives rise to a set of coupled nonlinear involving $f$ and $g$ and points to a supersymmetric system without hermiticity:
in other words, hermitian conjugation does not relate the supercharges. A similar analytic assumption has been made in the literature in connection with quantized systems and presents no difficulty in developing a consistent theoretical framework \cite{zno}. Below we present a couple of waveguide examples that are periodic and 
exhibit $\mathcal{PT}$-symmetry.

\subsection{The distribution $v_{}(z)= v_0 e^{i \beta z}$}
Following \cite{Lin,Panigrahi,Joglekar} we understand that the $\mathcal{PT}$-symmetry in optics focuses on the condition $n(z)=n^*(-z)$ for the complex refractive indices. The real part of the refractive index shows the peak index contrast while the imaginary part represents a loss or gain depending on its accompanying sign ($+,-$ respectively). In our current work we intend to explore its importance in the context of supersymmetry. As a simple choice, we take 
\begin{align} \label{n(z)v(z)}
    n (z)= n_0\Big(1+v(z)\Big)
\end{align}
where $v(z)$ acts as a small perturbation in the refractive index from the background material i.e.
$\lvert v(z) \rvert \ll 1$. If we choose $\mathcal{PT}$-symmetric choice for $v(z)$, a plane wave with amplitude $v_0$ namely,
$v (z)= v_0 e^{i \beta z}$, then for this class of $v(z)$, we have
on using (\ref{n(z)v(z)}) and (\ref{n2=W2W'}) corresponding to the upper sign, the following relationships

\begin{eqnarray}
 &&   n^{(+)} (z)= n_0(1+v_0 e^{i \beta z}) \label{n+1 v1} \label{n+} \\
&& {k^2}\Big(1 +2v_0  e^{i \beta z} + v_0^2 e^{i 2\beta z} \Big)=\varepsilon  - \lambda - (W^2 + W') \quad \label{n v1index and W}
\end{eqnarray}

We now proceed to determine the associated superpotential $W$ in the form of (\ref{W(z)fg}) which gives from (\ref{n v1index and W}) on equating the real and imaginary parts the coupled relations
\begin{subequations}
\begin{align}
     k^2 \Big(1 +2v_0 \cos{\beta z} +  v_0^2\cos{ 2\beta z} \Big) = \varepsilon -(f^2-g^2+f' + \lambda)  \\
     k^2 \Big(2 v_0 \sin{\beta z} + v_0^2 \sin{ 2\beta z} \Big) = -(2 f g +g')
\end{align}
\end{subequations}
These lead to the following matching conditions
\begin{align}
    \beta = \pm 2k, \quad g= \beta \frac{v_0}{2} \cos{\beta z} \quad  f=- \beta \frac{v_0}{2} \sin{\beta z} 
\end{align}
with $k = \pm \sqrt{\varepsilon - \lambda}$ and implies for the following representation of the superpotential
\begin{align} \label{v1in W form}
W = - \frac{\beta v_0}{2} \Big( \sin{\beta z}- i \cos{\beta z} \Big)
\end{align}

\begin{figure}[H]
\centering
\includegraphics[scale=0.5]{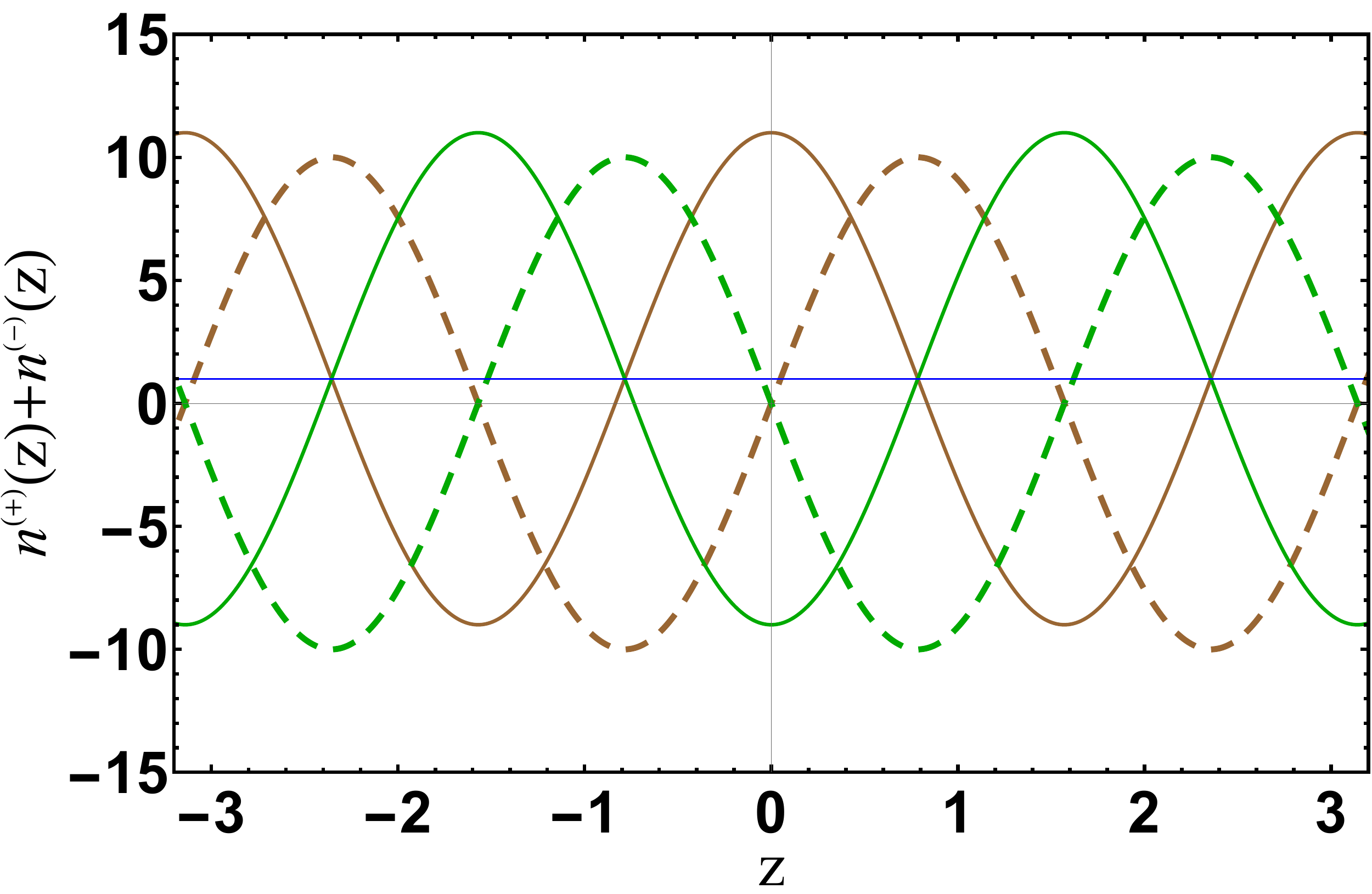}
\hfill
\caption{$n^{(+)}(z)+n^{(-)}(z)$ vs $z$ has been plotted. The brown lines show $n^{(+)}(z)$ and the green lines show $n^{(-)}(z)$. The solid lines represent the real part, and the dashed line represents the imaginary part. The blue line is the superposed profile of the two superpartner refractive indices. Here $v_0 = 10$ and $\beta = 2$}
 \label{plotting(n+ + n-)}
\end{figure}

Using (\ref{Vto W}) and (\ref{v1in W form}), we thus arrive at the form

\begin{equation}   \label{V-cos sin}
    V^{(-)}=\gamma \Big(1-(cos{2 \beta z}-\frac{2}{v_0} \cos{\beta z}) \\
    - i (\sin{2 \beta z}-\frac{2}{v_0} \sin{\beta z}) \Big) + \lambda
\end{equation}
where $\gamma=\frac{\beta^2v_0^2}{4}$, as a supersymmetric partner to $V^{(+)}$. The corresponding partner distribution of the refractive index reads $ n^{(-)} (z)= n_0(1-v_0 e^{i \beta z})$ in contrast to \eqref{n+}. An interesting off-shoot is that the supersymmetric partner indices add up to constant
\begin{align}  \label{n-1 v1}
   n^{(+)} + n^{(-)} = 2 n_0
\end{align}
That the partner complex refractive indices sum to a constant is the reflection of the fact that $SUSY$ is at work rather than any feature of GUP-based model, a situation much similar to the case of the behaviour of partner potentials of the harmonic oscillator. Of course, choice of the periodic distribution of $v(z)$ has been crucial in this case. In this context we observe that the supersymmetric partner of the refractive index $n^{(+)}$, namely 
$n^{(-)}$ has the negative sign in the perturbation $v(z)$. This represents the corresponding lossy nature of the partner refractive index, if we consider $n^{(+)}$ to be gainy one. This gives a naturally convenient explanation of a balanced composite loss-gain media. In \cite{miri2}, the relative permittivity distribution of the superpartner waveguide for a few profiles was identified. However, the basic analytical forms obtained in the present work for the partner potentials in the light of the complex splitting of the superpotential, along with the observation that the sum of the refractive indices corresponding to the supersymmetrically related  distributions turning out to be a constant, are new.  In the Figure \ref{plotting(n+ + n-)}  the individual variation of each refractive index is separately shown.

\begin{figure}[H]
\centering
\begin{subfigure}{0.45\textwidth}  \includegraphics[width=0.95\linewidth]{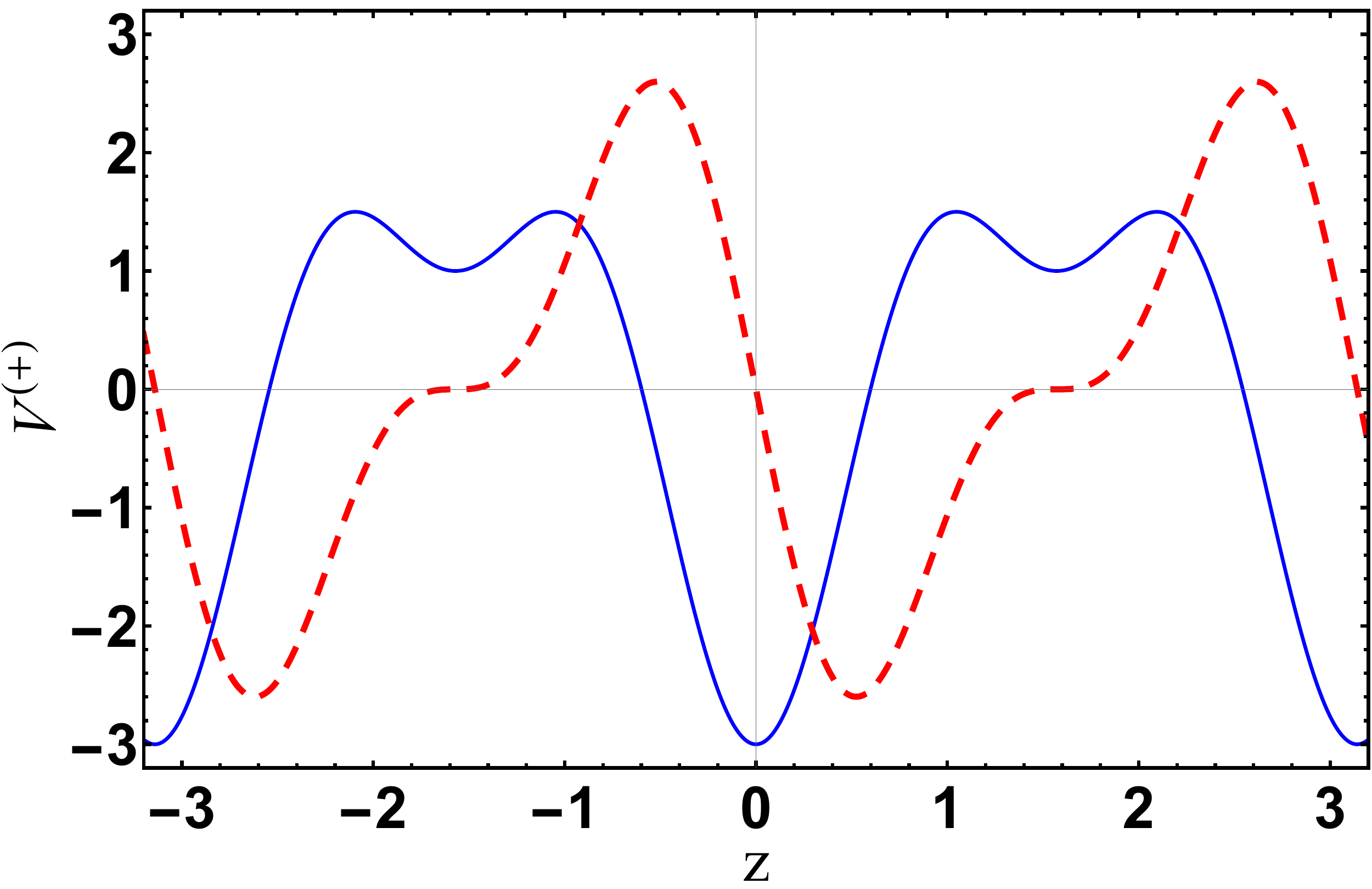}
    \caption{$V_{}^{(+)}$ vs $z$}
    \label{fig1:first}
\end{subfigure}
\hfill
\begin{subfigure}{0.45\textwidth}
    \includegraphics[width=0.95\linewidth]{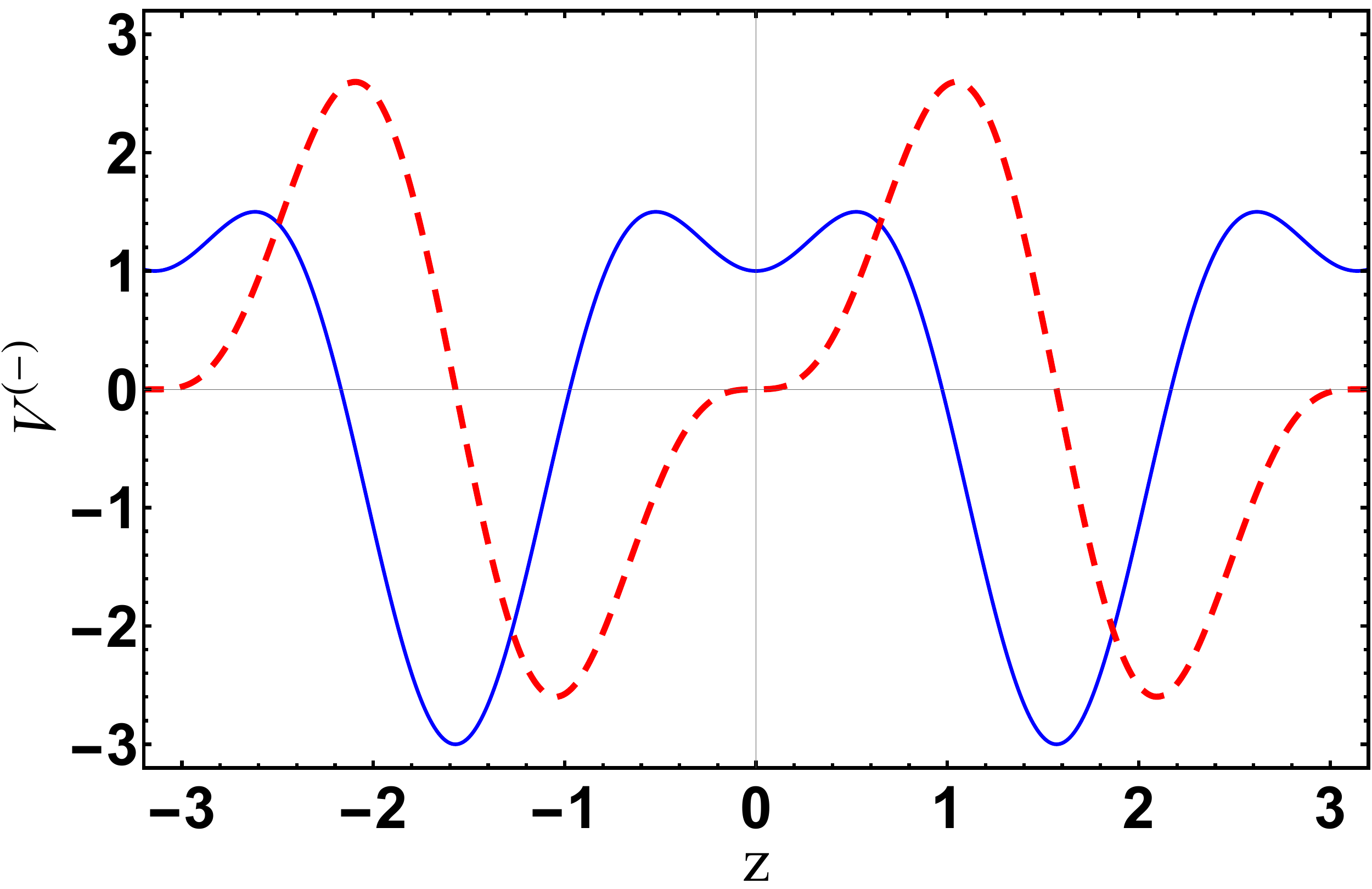}
    \caption{$V_{}^{(-)}$ vs $z$}
    \label{fig1:second}
\end{subfigure}
\hfill
\caption{Plots of $V_{}^{\pm}$ vs $z$ where the solid blue line represents the real part of the potential and the dashed red line represents the respective imaginary part. Here $k=1$, $\lambda=0$, and $v_0=1$.}
 \label{V1 plot}
\end{figure}

In Figure \ref{V1 plot} we display a simple computation of the supersymmetric partner potentials. Specifically, we observe in Figure \ref{fig1:first} that, while the real part of the original structure of $V^{(+)}$ depicts a repeated character of the multiple-well potential with narrow symmetrical drops, the shape of the imaginary part is also symmetrical in nature but passes through the origin. On the other hand, in Figure \ref{fig1:second}, for the plot of the partner $V^{(-)}$, the symmetrical nature of the multiple-well persists but, because of its mixed structure, there is a pronounced shift towards the positive $z$-axis in both the real part and  the imaginary part.   

In terms of recent interesting studies in unidirectional invisibility, the importance of $\mathcal{PT}$-symmetric complex refractive indices is noted. The
phase difference between the real and imaginary parts of the refractive index
controls the unidirectional reflectivity of the system \cite{Lin}. Attempts to create such novel invisibility materials using artificially engineered negative index metamaterials is an active field \cite{Schu, Alu, Cai, Smol, Ergin}.

\subsection{The distribution $v (z)= \nu_1 \cos 2\beta z + i \nu_2 \sin 2\beta z$}

This type of $v (z)$ follows from the $\mathcal{PT}$-symmetric periodic structure of the refractive index,  $n^{(+)}(z)=\eta_0+ \eta_1 \cos{2 \beta z}+ i \eta_2 \sin{2 \beta z}$, where $\eta_1$ corresponds to the peak real index contrast and $\eta_2$ represents the gain and loss of the
distribution. This form was recently advanced by Lin et al \cite{lin} to analyse the amplitudes of the forward and backward propagating waves outside of the grating domain and subsequently to acquire knowledge of the transmission and reflection coefficients. The purpose was to establish that $\mathcal{PT}$-symmetric periodic structures can act as unidirectional invisible media. Further, as a $\mathcal{PT}$-symmetric sinusoidal potential a similar form was studied in \cite{jones3} as an acting potential to deal with the beam propagation in $\mathcal{PT}$-symmetric optical lattices while the phase transitions of the eigenvalues were earlier investigated in \cite{roy}.
\begin{figure}[H] 
\centering
\begin{subfigure}{0.45\textwidth}  \includegraphics[width=0.95\linewidth]{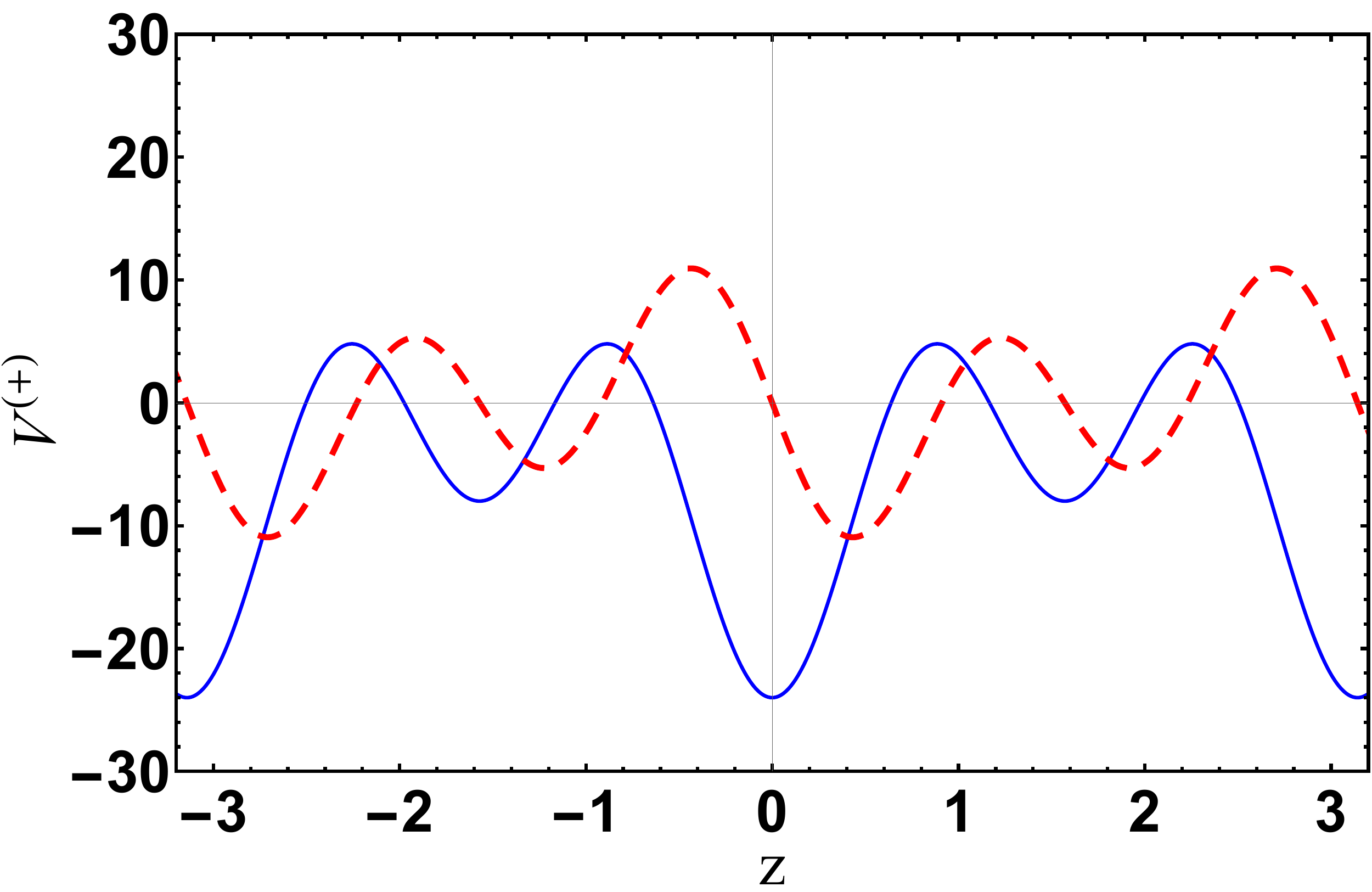}
    \caption{$V_{}^{(+)}$ vs $z$}
    \label{fig2:first}
\end{subfigure}
\hfill
\begin{subfigure}{0.45\textwidth}
    \includegraphics[width=\linewidth]{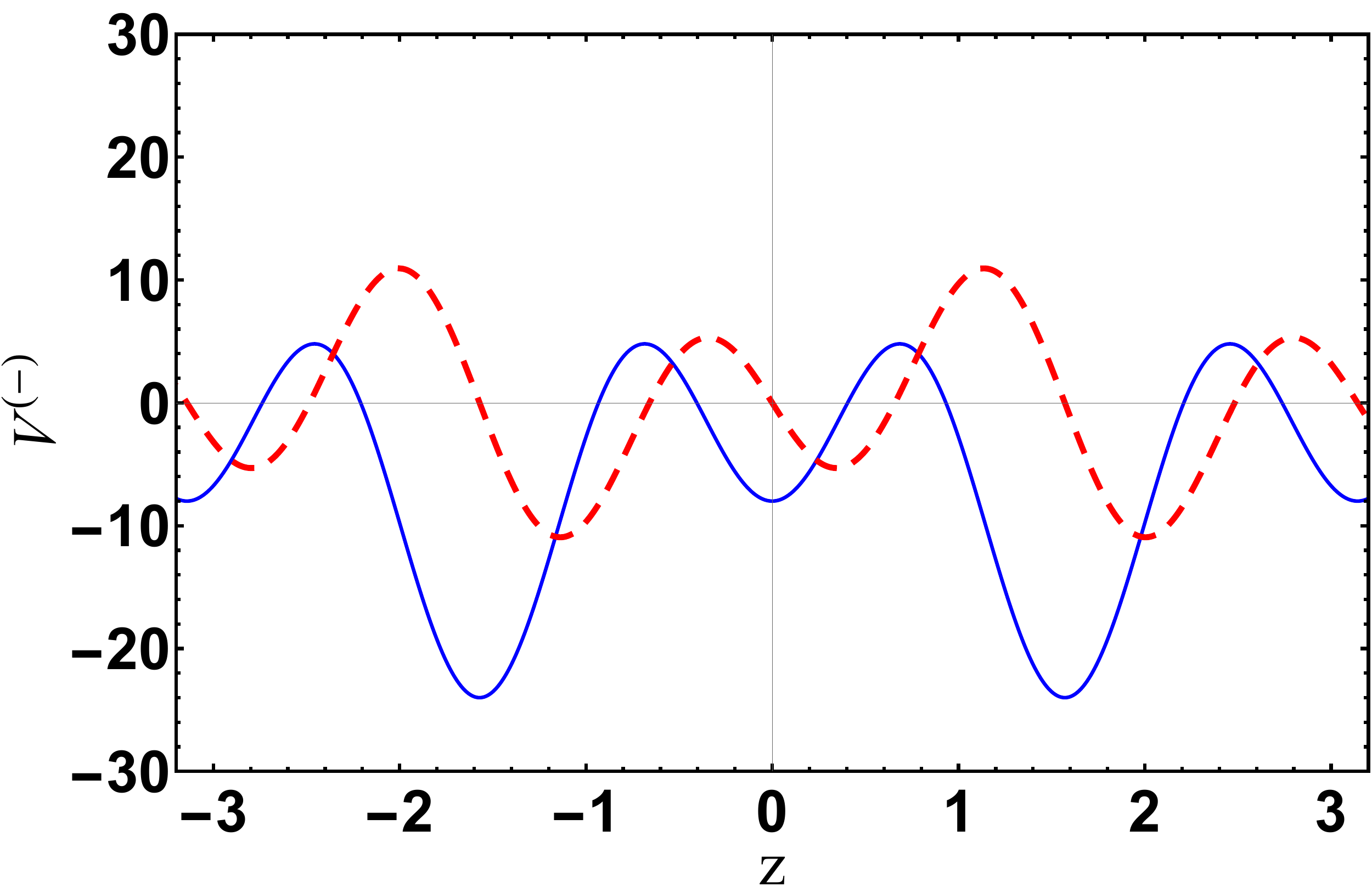}
    \caption{$V_{}^{(-)}$ vs $z$}
    \label{fig2:second}
\end{subfigure}
\hfill
\caption{Plots of $V_{}^{\pm}$ vs $z$ where the solid blue line represents the real part of the potential and the dashed red line represents the respective imaginary part. Here $k=1$, $\lambda=0$, $\nu_1=4$, and $\nu_2=2$.}
 \label{V2 plot}
\end{figure}

Given the above distribution of the refractive index, we can utilize (\ref{n2=W2W'}) to obtain 

\begin{align}  \label{n v2index and W}
    k^2 \Big( 1 +\nu_1 \cos{2\beta z} +i\nu_2 \sin{2\beta z}\Big)^2=\varepsilon -\lambda - (W^2 -W')
\end{align}
where $\nu_1 = \frac{\eta_1}{\eta_0}$ and $\nu_2 = \frac{\eta_2}{\eta_0}$. Then from (\ref{n v2index and W}) and (\ref{W(z)fg}), the corresponding real and imaginary part leads respectively to 
\begin{subequations}
\begin{equation}
    k^2 \Big(1 +\nu_1^2 \cos^2{2\beta z} -\nu_2^2 \sin^2{2\beta z}\\
    +2\nu_1 \cos{2\beta z} \Big) = \varepsilon -f^2+g^2-f'-\lambda 
    \end{equation}
    \begin{equation}
         k^2 \Big(2 \nu_1 \nu_2 \cos{2\beta z} \sin{2\beta z} +2 \nu_2 \sin{2\beta z} \Big) = -2 f g -g'
    \end{equation}
   \end{subequations}
On inspection, the following set of solutions emerges 
\begin{equation}
    \beta=\pm k, \quad g= \nu_2 \beta \cos{2 \beta z}, \quad  f=- \nu_1 \beta \sin{2\beta z} 
\end{equation}
along with

\begin{equation}
 \varepsilon-\lambda= k^2(1+\nu_1^2-\nu_2^2)
\end{equation}
and the superpartner wave-guide index acquires the form
\begin{align}
    n^{(-)}(z)= \eta_0\Big( 1-\nu_1 \cos{2 \beta z} - i \nu_2  \sin{2\beta z} \Big)
\end{align}
Here too, the two index profiles satisfy the same constraint as in (\ref{n-1 v1}). The profiles of the corresponding $V_{}^{(\pm)}$ are sketched in Figure \ref{V2 plot}. In contrast to the real part the imaginary part distinctly reveals a sinusoidal character with varying amplitudes. 

\section{Summary}

 We subjected the SHE to a supersymmetric treatment within a $\mathcal{PT}$ optical structure in which the underlying refractive index distribution has a longitudinal variation. The superpartner of the index profile was analytically evaluated and closed form solutions of some typical distributions were obtained by solving a pair of coupled equation involving the real and imaginary components of the superpotential. The features of the index distribution corresponding to the supersymmetric partners were graphically illustrated. We have highlighted that our $\mathcal{PT}$-symmetric refractive indices and their corresponding supersymmetric partner potentials result in a composite balanced loss-gain behaviour.

\section{Acknowledgements}
BB is grateful to Brainware University for infrastructural support. SS thanks the Shiv Nadar IoE (deemed University) for financial assistance in the form of senior research fellowship. RG acknowledges IIT Delhi for financial support.

\section{Data availability statement}

All data supporting the findings of this study are included in the article.

\end{document}